\documentclass[review,5p]{elsarticle}

\usepackage{hyperref}
\usepackage[latin1]{inputenc}
\usepackage{color}           
\usepackage{soul}            
\usepackage{amssymb}
\usepackage{amsfonts}
\usepackage{amsmath}
\usepackage{wasysym}
\usepackage{graphicx}
\usepackage{epsfig}
\usepackage{amsthm}
\usepackage{bm}

\journal{Journal of \LaTeX\ Templates}

\begin{document}

\begin{frontmatter}

\title{Recycling the resource:  Sequential usage of shared state in quantum teleportation with weak measurements}

\author{Saptarshi Roy\(^1\), Anindita Bera\(^{1,2,3}\), Shiladitya Mal\(^1\), Aditi Sen(De)\(^1\), Ujjwal Sen\(^1\)}
\address{\(^1\)Harish-Chandra Research Institute, HBNI,  Chhatnag Road, Jhunsi, Allahabad 211 019, India\\
\(^2\)Department of Applied Mathematics, University of Calcutta, 92 Acharya Prafulla Chandra Road, Kolkata 700 009, India\\
\(^3\)Racah Institute of Physics, The Hebrew University of Jerusalem, Jerusalem 91 904, Givat Ram, Israel}




\begin{abstract}
Complete measurements, while providing maximal information gain, results in destruction of the shared entanglement. In the standard teleportation scheme,  the sender's measurement on the shared entangled state between the sender and the  receiver has that consequence. We propose here a teleportation scheme involving weak measurements which can sustain entanglement upto a certain level so that the reusability  of the shared resource  state for reattempting teleportation  is possible. The measurements are chosen in such a way that it is weak enough to retain entanglement and hence can be reused for quantum tasks, yet adequately strong to ensure quantum advantage in the protocol. In this scenario, we report that at most six sender-receiver duos can reuse the state for reattempting teleportation, when the initial shared state is entangled in a finite neighborhood of the maximally entangled state and for a suitable choice of  weak measurements. However, we observe that the reattempt number decreases with the decrease in the  entanglement of the initial shared state. Among the weakening strategies studied, Bell measurement admixed with white noise performs better than any other low-rank weak measurements in this situation. 
\end{abstract}

\begin{keyword}
\texttt{Quantum information}\sep \texttt{Quantum measurements}\sep \texttt{Quantum teleportation}
\end{keyword}

\end{frontmatter}

\section{Introduction}
Over the last decade, it has been established that 
 next generation  communication technology can be
revolutionised by employing laws of quantum theory. In this regard, the invention of quantum
teleportation \cite{Bennett_1993} has played a crucial role in the advancement of quantum
communication. It is a protocol by which one can send the information in a quantum state to a remote party
without sending the system itself physically. 
After the initial proposal, several attempts have been made to generalize it, which  include construction of the optimal teleportation protocol for a general resource state \cite{frank-ver},  characterization of its performance  via both fidelity and its deviation \cite{HorodeckiPRA98, dag2018}, understanding the relation between entanglement of the resource state and the  fidelity \cite{frank-ver}, extension to continuous variable systems \cite{telecont, tele_contvar_exp},  teleportation in a multiparty setting \cite{Kimble2008, Murao98}. 
On the other hand, based on the teleportation protocol, many other quantum tasks like quantum repeaters \cite{Cirac_1998}, quantum gate teleportation
\cite{Chuang_1999}, measurement-based computing \cite{Briegel_2001}  have been designed which facilitate the progress of quantum
information and communication.  
Relaizations of these tasks in various  physical systems, like photonic qubits \cite{tele_review, tele_photon_exp, tele_photon_exp2, photon_loss_distance}, nuclear magnetic resonance
\cite{Nielsen_1998}, trapped ions \cite{Barrett_2004, tele_ion_exp} play a key role in the developments of quantum communication (cf. \cite{tele_exp_coldatom, tele_supercond, tele_exp_light_matter}). Recently, long-distance teleportation using photons has also been achieved between two cities which are at a distance of a thousand kilometers
\cite{tele_photon_exp2}.

Information  cannot, in general, be gained through quantum measurement without disturbing the  system \cite{disturbance}.
 Typically, in a teleportation protocol,  a shared state  acting as a channel can be used only once if the task is performed employing a complete projective measurement. This is due to the fact that in this scenario, quantum correlations present in the channel or in the resource state between the sender and the receiver is completely destroyed after the measurement.  In this respect, one can ask the following question:  \emph{If the sender does not perform a (complete) projective measurement, can the resource state remain useful  by saving  part of its entanglement content, for possible utilization in the future round?}
 To address it, 
non-projective measurements or weak or unsharp measurements \cite{unsharp} can be carried out,  which disturb the state less  at the cost of a reduced information gain, thereby creating a trade-off between measurement disturbance and information gain.   Note that although  the joint measurement and the channel are both  key elements in  teleportation, the role of the former   is less studied than that of the latter  on the performance of the protocol \cite{Gisin_2019}.

 In this work,  we investigate  different weak measurement strategies, which are a class of  positive operator-valued measurements (POVMs) to achieve maximal number of recycles of a fixed teleportation channel, with the maximal number being referred to as the maximal reattempt number (MRN). Such number is computed by maintaining the teleportation fidelity  beyond the classical one at each round. 
Specifically, when  the standard Bell measurement is weakened by admixing  it with  white noise and for a   shared maximally entangled state as a resource, 
we report that the channel can be used at most six times, while still attaining quantum advantage in  the  teleportation protocol. This kind of weakening is known to optimize information gain-disturbance in case of two outcome measurements \cite{tradeoff}. It is interesting to note that in case of violation of Bell inequalities by multiple observers on one side, at most two observers can violate the inequality \cite{shilu-bell,shilu-bell2}, and in case of witnessing entanglement in the same scenario,  at most twelve observers can detect bipartite entanglement, with 
another observer situated at a distant location \cite{aniweak}. Therefore, the results  here indicate that the multiple-round teleportation fidelity with weak measurements  has apparently an intermediate standing between Bell inequality violation
 and sharing of entanglement. This is in sharp contrast to the situation while using complete projective measurements, as then, entanglement and teleportation fidelity vanish together for two-qubit states, while Bell inequality violation is absent in a larger class. See \cite{whatisnonlocality} in this respect.
 We also find that the number, six, remains unaltered even for non-maximally entangled states having entanglement beyond a certain critical value.   
 For a fixed measurement scheme, we observe that MRN  decreases with the decrease  of entanglement of the shared state at each round although  a plateau with respect to the content of entanglement  is found  for  a fixed value of MRN. 
 Moreover, we extend our study to other prototypical weak measurements. For example,  a specific Bell state  smeared by mixing states from its support or from orthogonal support leads to a  lower value of  MRN, thereby a weaker value of the corresponding entanglement, compared  to the case of Bell measurement with white noise. 
 
Let us now illustrate how our protocol can be useful in a realistic scenario. Suppose a broker in a stock market wants to send information to one of her/his clients about some investment via one of her/his employees. Although the stock prices are public, the employee is supposed to remain unaware of the investment strategy the broker wants to send his client. We now generalize the situation to the case where the strategy involves quantum information (qubits) to be sent from the broker to her/his client. The employee and the client share a quantum channel, possibly a maximally entangled state, and the information is sent to the client by employing a teleportation scheme. The stock market is obviously fluctuating and the broker wants to keep the flexibility of changing her/his information (the qubit that has to be teleported) depending on the market status without any additional resources (entangled channels) shared by the employee and the client. In such a situation, our scheme of reattempting teleportation over multiple rounds via recycling the same resource state becomes particularly useful. We want to stress here that such an example is illustrative, and our scheme would be useful in all such situations where temporal flexibility in terms of when information needs to be sent has to be taken care of via allowing the possibility of recycling the resource. 
 
To avoid any confusion, we want to mention here that imperfect teleportation has been studied thoroughly both from the point of view of noisy resource states and faulty measurement schemes. Error sources are typically beyond the control  of the experimentalist and their effects are studied from their detrimental effect on the performance of the teleportation scheme. 
In this work, we adopt a completely different motivation of recycling the resource for implementing quantum teleportation over multiple attempts. 
We achieve this aim by replacing the usual Bell measurements by a suitable POVM (weak measurement). Note that, here the apparently
\emph{imperfect measurement} is a conscious choice in the protocol-design which enables recycling of the resource. 
So, the POVMs in our case are not some uncontrollable detrimental evil but rather a strategy for \emph{recycling the resource}.


This paper is organized in the following way. In Sec. \ref{sec:formalism}, we first describe  the  standard teleportation protocol with projective measurements, and then introduce the concept of teleportation via weak measurements which helps us to answer the main question of recycling the resource state.   With different weak measurement strategies, we present the analysis of reattempt when the resource state is maximally entangled in  Sec. \ref{sec:maxent},  while investigation  for other non-maximally entangled resource state is carried out in Sec. \ref{sec:nonmax}. We summarize our results in Sec. \ref{sec:conclusion}. 

\section{Methodology: Reusing the Resource}
\label{sec:formalism}
Before introducing the protocol which enables reattempting teleportation of the resource, let us first  briefly discuss  the standard teleportation protocol, which in turn guide us to the construction of the new one.  Let us consider a state, $\rho^{AB}$,  shared between two parties, say, Alice ($A$) and Bob ($B$). Alice wants to teleport an arbitrary single qubit state,
\begin{equation}
|\eta\rangle^{A'} = a|0\rangle+ b|1\rangle,
\label{eq:unknown_state}
\end{equation}
with $|a|^2 + |b|^2 = 1$ to Bob. The initial state then reads as
\begin{equation}
\rho^{A'AB}=|\eta\rangle^{A^{\prime}}\langle\eta|\otimes\rho^{AB},
\label{eq:totalstate}
\end{equation} 
In this scenario, $A$ performs a joint (complete) projective measurement on the $A'A$ part of $\rho^{A'AB}$ and communicates classically the measurement results to $B$ who acts locally according to $A$'s communication to reproduce the input.  The objective of the protocol is to maximize the fidelity, $f$, where maximization is performed over all operations allowed in the protocol by $A$ and $B$, averaged over all input states to be teleported to $B$ and is given by
\begin{equation}
f = \int \langle\eta|\rho^B|\eta\rangle ~d\eta,
\label{eq:fidelity}
\end{equation}
where $\rho^B$ is the reduced density matrix of $B$ after all operations performed by $A$ and  $B$. When the resource state is a general two qubit mixed state, the optimal teleportation protocol and the corresponding fidelity are derived in Refs. \cite{HorodeckiPRA98, frank-ver}.  Specifically,  if the resource state is Bell diagonal \cite{bd}, the teleportation protocol, involving Bell basis measurements in $A'A$ and Pauli rotations at $B$, yields the optimal fidelity. It is also known that for an arbitrary  shared separable state, the fidelity is \(2/3\), which we refer as the classical one. Note  here that the attainment of the maximal fidelity would result in the complete destruction of the resource (entanglement) between $A$ and $B$. Therefore, in this communication scheme, there is \emph{no possibility of reusing the resource in the teleportation channel, i.e., the shared state.}
 
 \subsection{Prescription for reusing resource state with  weak measurements}
 \label{weak-me}
 
In this paper, we consider a shift in ``paradigm" for teleportation protocols, from ones which strive to maximize the fidelity, to those whose objective is to \emph{reattempt teleportation for maximal number of rounds}, while ensuring the fidelity at each round to be nonclassical \cite{massar}, i.e., $f > 2/3$.
 We demonstrate that Alice can achieve this by shifting from the projective measurement strategy to one that employs weak (unsharp) measurements \cite{unsharp}. Such  measurements can be described by a  set of POVMs, $\{\mathcal{M}_i\}$, such that 
\begin{equation}
\sum_i \mathcal{M}_i = \mathbb{I} \text{  and  }  \mathcal{M}_i  \geq  0 \,\, \forall i.
\end{equation}
\begin{figure}
\includegraphics[width=8.5cm]{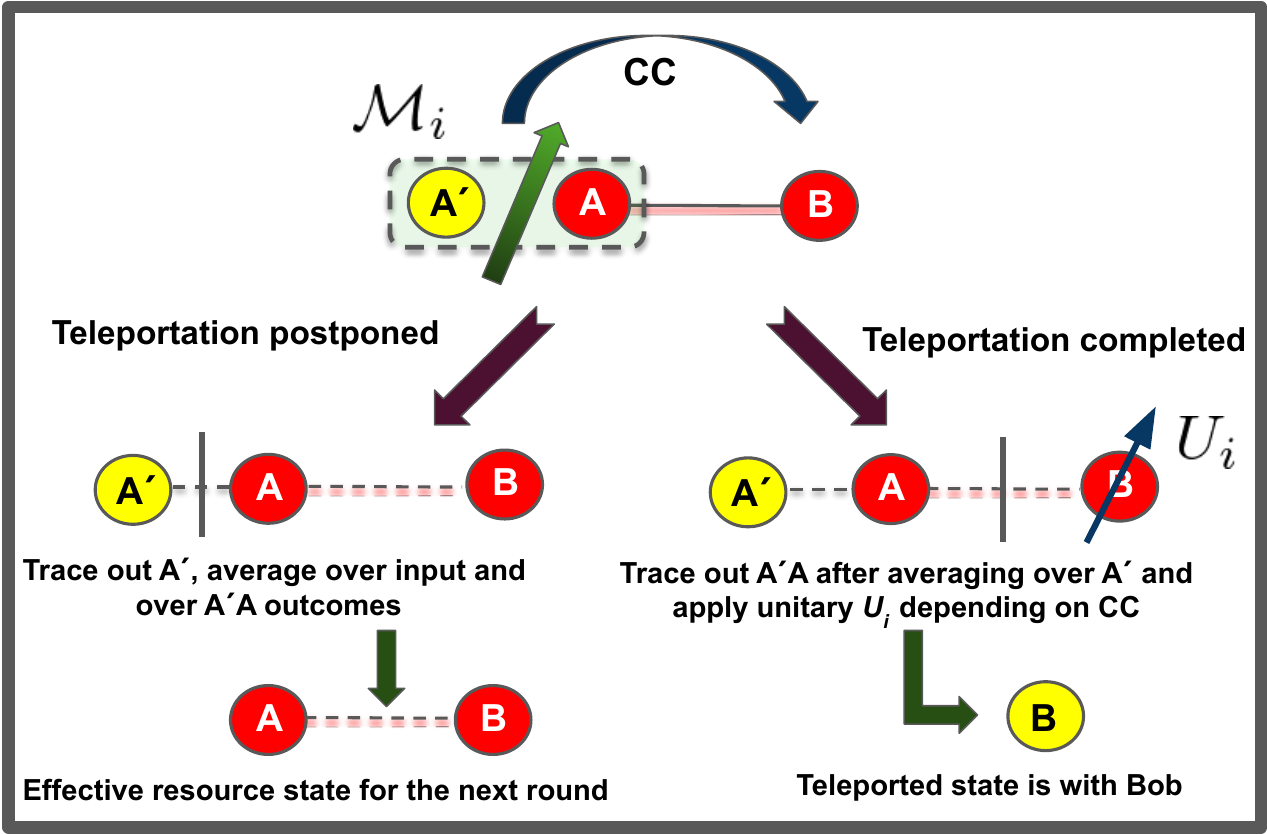}
\caption{Schematic representation of the teleportation protocol which allows  reusability of the resource state. CC denotes the classical communication. \(\mathcal{M}_i\) and \(U_i\) are respectively the set of  positive operator-valued measurements  on $A'A$  and the unitary operators at $B$'s port.}
\label{fig:protocol}
\end{figure}
After the weak joint measurement by A, the measurement results are classically communicated to B, who now has two choices (see Fig. \ref{fig:protocol}): 
\begin{enumerate}
\item $B$ can complete the teleportation process by applying appropriate unitary operators depending on the measurement outcome.

\item $B$ does nothing and leaves the protocol, allowing for a new pair of $A$ and $B$ to reuse the remaining  resource (if any) to teleport in the next round at some later time.
\end{enumerate}

\noindent If $B$ declines to complete the teleportation in the first round, the new $B$ has exactly the same  two options after the measurement in the second round. Similar situation occurs for all successive rounds. Note, however, that this is not an indefinite process. The finiteness of the maximal number of rounds is induced by the fact that the completion of the protocol at any given intermediate round would have to yield a nonclassical teleportation fidelity, thereby requiring an entangled state shared by $A$ and $B$. Therefore, finding the maximal number of reusability of a given channel is one of the main goals of our work. We call this number as the \emph{maximal reattempt number}. It is a function of the teleportation protocol employed, including the shared state, the measurements at Alice's lab and the unitary operations  at Bob's lab.


Nevertheless, the post measurement state when the i-\text{th} POVM element, $\mathcal{M}_i$, clicks can be expressed as 
\begin{equation}
\rho_i^{A'AB} =\frac{1}{\mathcal{N}} \Big(\sqrt{\mathcal{M}_i^{A'A}}\otimes \mathbb{I}_2^{B}\Big) \rho^{A'AB} \Big(\sqrt{(\mathcal{M}^{A'A}_i)^{\dagger}} \otimes \mathbb{I}_2^{B}\Big), 
\end{equation}
where $\mathcal{N}$ is the normalization constant, which is simply the probability, $p_i$, of the i-th outcome, given by 
\begin{equation}
 p_i =\text{Tr} \Big[ \Big(\sqrt{\mathcal{M}^{A'A}_i}\otimes \mathbb{I}^{B}_2\Big) \rho^{A'AB} \Big(\sqrt{(\mathcal{M}^{A'A}_i)^{\dagger}}\otimes \mathbb{I}_2^{B}\Big)\Big]. 
\end{equation}
If $B$ wants to finish the process, depending on the measurement outcome (which $A$ communicates to $B$), the state of B, rotated by appropriate unitaries, \(\{U_i^B\}\),  as well as averaged over all possible inputs and measurement outcomes  can be represented as
\begin{equation}
\rho^B = \sum_i p_i \hspace{0.2cm} U_i^B \Big( \text{Tr}_{A'A} \hspace{0.2cm}  \rho_i^{A'AB}  \Big)  (U_i^B)^\dagger,
\label{eq:PMS_bob}
\end{equation}
and the corresponding fidelity is given in Eq.~\eqref{eq:fidelity}.
If $B$ refuses to complete the protocol, the effective state shared between $A$  and $B$ for subsequent rounds is obtained by tracing out $A'$ after averaging over the measurement outcomes and the input state, $|\eta\rangle$ (see Eqs.~\eqref{eq:unknown_state} and \eqref{eq:totalstate}), which reads as
\begin{eqnarray}
\rho^{AB}_{\text{effective}} =\text{Tr}_{A'} \int \sum_i \Big(\sqrt{\mathcal{M}_i^{A'A}}\otimes\mathbb{I}_2^B \Big) \rho^{A'AB} \nonumber\\ \Big(\sqrt{(\mathcal{M}_i^{A'A})^{\dagger}}\otimes \mathbb{I}_2^B\Big) d\eta.  
\label{eq:rho-effective-gen}
\end{eqnarray}
For a given shared state, 
the above steps can be repeated to obtain MRN for a fixed value of $f$ provided the fidelity in each round is beyond the classical limit.

\section{Maximally entangled state as the initial resource}
\label{sec:maxent}

In this section, we focus on the maximally entangled state shared between Alice and Bob, given by
\begin{equation}
|B_1^{AB} \rangle = \frac{1}{\sqrt{2}}\big(|00\rangle + |11\rangle\big),
\end{equation}
as the initial resource to compute MRN for different choices of weak measurement strategies.  \(|B_1\rangle\) is one of the triplets in the Bell basis \cite{Bell-basis} and the results obtained by using this state will be same for any other shared maximally entangled states.  

\subsection{Weakening Bell measurements via depolarization}

Instead of performing Bell basis measurement, Alice applies POVMs, \(\{\mathcal{M}_i^{A'A}\}_{i=1}^4\),   on the $A'A$ part which is formed by mixing Bell measurements  with a completely depolarizing (maximally mixed) state. The i-th element of POVM can be represented as 
\begin{equation}
\mathcal{M}_i^{A'A} = \lambda_1 |B_i^{A'A}\rangle\langle B_i^{A'A}| + \frac{1-\lambda_1}{4}\mathbb{I}_4^{A'A},
\label{eq:whitenoise}
\end{equation}
 where $\lambda_1 \in (0,1]$ is the sharpness parameter and $|B_i^{A'A}\rangle$ is one of the Bell states. Note that $\lambda_1=1$ corresponds to the projective measurement.
The  initial state  on which the measurement has to be performed is $|\psi^{A'AB}\rangle = |\eta\rangle^{A'} \otimes |B_1^{AB}\rangle$. 
Now we present a sketch of the Stinespring  dilation to implement the POVM considered above. First we consider a $4$-dimensional auxiliary system, $S$, defined by the basis $\lbrace |i^S \rangle \rbrace$, $i = 0, 1, 2, 3$ with $|\psi^{A'AB}\rangle$. With the initial state of the ancillary system being $|0^S\rangle$, we then have the state $|0^S\rangle \otimes |\psi^{A'AB}\rangle$. Now we couple the system and the auxiliary using the unitary $U^{SA'A}\otimes \mathbb{I}^B$ such that:
\begin{eqnarray}
U^{SA'A}\otimes \mathbb{I}^B |0^S\rangle &\otimes& |\psi^{A'AB}\rangle = \nonumber \\
 \sum_{i=0}^3 |i^S\rangle &\otimes& \big( \sqrt{\mathcal{M}_i^{A'A}} \otimes \mathbb{I} \big) |\psi^{A'AB}\rangle.
\end{eqnarray} 
The existence of the unitary $U^{SA'A}$ is guaranteed by the following conditions
\begin{eqnarray}
\sum_i \mathcal{M}_i^{A'A} = \mathbb{I}, \hspace{0.2cm} \mathcal{M}_i^{A'A} \geq 0, \text{ and } (\mathcal{M}_i^{A'A})^\dagger = \mathcal{M}_i^{A'A}. \nonumber \\
\end{eqnarray}
Now, a projective measurement in the $\lbrace |i^S \rangle \rbrace$ basis yields the desired outputs.

Therefore following the strategy developed in Sec. \ref{sec:formalism}, $B$ might wish to complete the teleportation process and acts his qubit with Pauli operators just as in the standard teleportation scheme. The resulting post measurement state with $B$ is then given by
\begin{eqnarray}
 \rho^{B} &=& 
\begin{bmatrix}
   |a|^2\lambda_1 + \frac{1-\lambda_1}{2}      & ab^* \lambda_1 \\
      ba^* \lambda_1     &  |a|^2\lambda_1 + \frac{1-\lambda_1}{2}
\end{bmatrix} \nonumber \\  &=& \lambda_1 |\eta\rangle\langle\eta| + \frac{1-\lambda_1}{2}\mathbb{I}_2.
\label{eq:bobstate}
\end{eqnarray}
 Hence the corresponding  fidelity, following Eq. \eqref{eq:fidelity},  reads as
\begin{equation}
f(1,\lambda_1) = \int  \langle\eta|\rho^B|\eta\rangle \hspace{0.15cm} d\eta = \frac{1+\lambda_1}{2},
\label{eq:f1}
\end{equation} 
where \(``1"\) and \(``\lambda_1"\) in the arguments refer to the initial maximally entangled state, which can be thought of as a Werner state, $p|B_1^{AB}\rangle\langle B_1^{AB}|+ \frac{1-p}{4}\mathbb{I}_4^{AB}$, with $p=1$, and the sharpness parameter of the POVM respectively. 
Note $f_1(1, \lambda) > \frac{2}{3}$ for $\lambda > \frac{1}{3}$.

On the other hand,  $B$ might not want to go on with the teleportation protocol at this round and therefore, does nothing. In this situation, the effective state shared between $A$ and $B$ is computed by performing averages over all the post-measurement states after the POVMs and over all possible input states, \(|\eta \rangle\). If the initial shared state is \( |B_1^{AB}\rangle \) and the POVMs are of the form \(\{\mathcal{M}_i^{A'A}\}\), the resulting state for the i-th outcome is the Werner state, given by
\begin{equation}
\rho^{AB}(1,\lambda_1) = p(\lambda_1)|B_1^{AB}\rangle\langle B_1^{AB}|+ \frac{1-p(\lambda_1)}{4}\mathbb{I}_4^{AB}, 
\label{eq:werner-phi}
\end{equation}
with
\begin{equation}
p(\lambda_1) = \frac{1}{2}\Big( 1- \lambda_1 + \sqrt{(1-\lambda_1)(1+3\lambda_1)} \Big).
\label{eq:p1}
\end{equation}
Here in \(\rho^{AB} ( 1, \lambda_1)\), arguments have the same meaning as in $f(1,\lambda_1)$. It is known that the state is entangled for $p > \frac{1}{3}$. Note that if $A$ and $B$ now use \(\rho^{AB}(1,\lambda_1)\) as the initial resource state and $A$ decides to perform the projective measurement, the maximal teleportation fidelity  after the second round turns out to be \(f(p(\lambda_1),\lambda_2=1) = (1 + p(\lambda_1))/2\), where $\lambda_2$ is the sharpness parameter of the second round measurement.   It  is greater than  $2/3$ when 
 $\lambda_1 < \frac{1}{3}(1+\sqrt{3}) \approx 0.9107$. Therefore, we find a range $\frac{1}{3} < \lambda_1 < \frac{1}{3}(1+\sqrt{3})$ for which both the  fidelities, $f(1, \lambda_1)$ and $f(p(\lambda_1),\lambda_2=1)$, obtained in the first round by weak measurement and second round by projective measurement  respectively are greater than the classical bound of $2/3$, thereby  \emph{ confirming the plausibility of reusing the shared resource state for reattempting teleportation in more than one round}.

\begin{figure}[ht]
\includegraphics[width=0.9\linewidth]{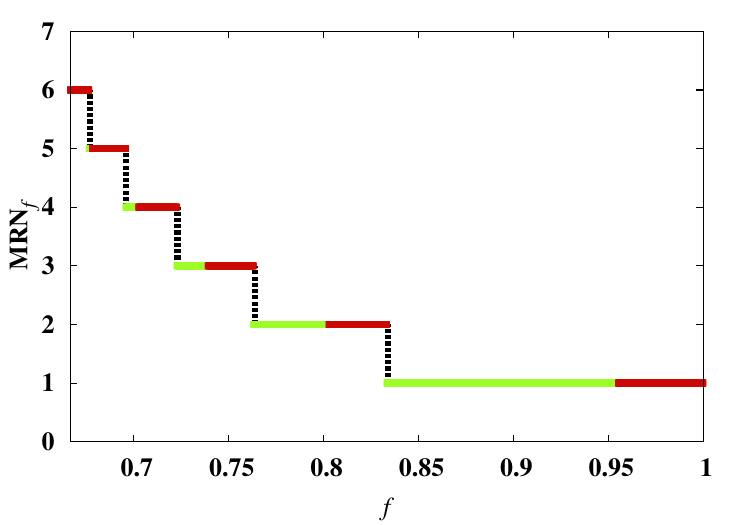}
\caption{MRN$_f$ vs fidelity ($f$). The initial resource state is a maximally entangled state. MRN$_f$ is computed for a  fixed value of the  fidelity, $f$, at each round, and for a fixed shared channel. The dark (red) lines indicate situations for which the effective state for the next round is unentangled, while, the grey (green) lines denote entangled effective states for the next round.  Both the axes are dimensionless. }
\label{fig:mrnvsf}
\end{figure}

The calculation of the average fidelity and the effective states can be performed generally for any subsequent rounds based on  three lemmas which we state next. 

{\it Lemma 1}:
{\it The teleportation fidelity of a Werner state of probability $p'$,  given in Eq. (\ref{eq:werner-phi}), when subjected to a weak Bell measurements of the type in Eq. \eqref{eq:whitenoise}, having the sharpness parameter $\lambda'$, at any round is given by
\begin{eqnarray}
f(p',\lambda') = \frac{1+p'\lambda'}{2}.
\label{eq:fid_eff}
\end{eqnarray}}
\begin{proof}
The Werner state of probability $p'$, shared between $A$ and $B$  reads as
\begin{eqnarray}
\rho^{AB}(p') = p' |B_1^{AB}\rangle\langle B_1^{AB}| + \frac{1-p'}{4}\mathbb{I}_4^{AB}.
\end{eqnarray}
When a weak Bell measurement of sharpness $\lambda'$ is employed, the maximally entangled part of the above eq. i.e., $|B_1^{AB}\rangle\langle B_1^{AB}|$ of $\rho^{AB}(p')$, yields a fidelity of $f(1,\lambda')=\frac{1+\lambda'}{2}$, while the maximally mixed part, $\frac{1}{4}\mathbb{I}_4^{AB} = \frac{1}{2}\mathbb{I}_2^{A} \otimes \frac{1}{2}\mathbb{I}_2^{B} $ of $\rho^{AB}(p')$, gives a fidelity of $1/2$, which is independent of the values of $\lambda^{\prime}$. The later simply follows from the fact that the state with Bob,  $\frac{1}{2}\mathbb{I}_2$, remains unaltered on any measurements in Alice's part. 
Finally, we obtain
\begin{eqnarray}
f(p',\lambda') &=& p' f(1,\lambda') + (1-p')\frac{1}{2} \nonumber \\
&=& p' \Big(\frac{1+\lambda'}{2}\Big) + \frac{1-p'}{2} \nonumber \\
 &=& \frac{1+p'\lambda'}{2},
\end{eqnarray}
and hence the proof.
\end{proof}


Let us now consider the case when a fixed value of fidelity, say $f(p',\lambda')$ would have to be achieved 
with a Werner state of probability $p'$ used as resource. In this case, the sharpness parameter $\lambda'$ has to be chosen in such a way so that
\begin{eqnarray}
f(p',\lambda') &&= \frac{1+p'\lambda'}{2}, \nonumber \\
\Rightarrow \lambda' &&= \frac{1}{p'}\Big(2 f(p',\lambda') -1 \Big).
\label{eq:lambdaf}
\end{eqnarray}
After evaluating the fidelity of a given Werner state, we want to compute the effective state for the next round of teleportation when we have a Werner state as resource and is subjected to weak Bell measurements, which we do in the following two lemmas.

{\it Lemma 2}:
{\it When a product state of the form
\begin{equation}
\tilde{\rho}^{AB} = \frac{1}{2}\mathbb{I}_2^{A} \otimes \Lambda^B,
\label{eq:product-resource}
\end{equation}
with $\Lambda^B$ being any arbitrary single qubit mixed state from $\mathbb{C}^2$, is used as a resource for teleportation with weak Bell measurements of the type in Eq. \eqref{eq:whitenoise}, the effective state for the next round remains the same as the initial  product resource.}
\begin{proof}
We begin the proof by noting down two important facts:
\begin{enumerate}
\item The Haar uniform average of all possible inputs is the maximally mixed state 
\begin{equation}
\int  |\eta\rangle \langle \eta| \vspace{1cm} \text{ } d \eta = \frac{1}{2}\mathbb{I}_2.
\label{eq:avg-state}
\end{equation}
\item Using linearity, we have
\begin{eqnarray}
 \sum_i \Big( E_i p\rho E_i^{\dagger} + E_i (1-p)\sigma E_i^{\dagger} \Big) \nonumber\\
= \sum_i E_i \Big(p\rho+(1-p)\sigma\Big) E_i^{\dagger}.
\label{eq:interchange}
\end{eqnarray}
Evolving various states via the same dynamical map and summing them is same as summing the states up with the given weight factors and then evolving with the same dynamical map. 
\end{enumerate}

Let us consider an initial resource state as in Eq. \eqref{eq:product-resource}, which after operating $\{\mathcal{M}_i^{A'A}\}$, the effective state becomes
\begin{eqnarray}
\tilde{\rho}^{AB}_{\text{eff}} =&&\text{Tr}_{A'} \int \sum_i \Big(\sqrt{\mathcal{M}_i^{A'A}}\otimes \mathbb{I}_2^B \Big) \times \nonumber\\ 
&&|\eta\rangle^{A^\prime} \langle\eta|\otimes\tilde{\rho}^{AB}  \Big(\sqrt{(\mathcal{M}_i^{A'A})^{\dagger}}\otimes \mathbb{I}_2^B\Big)  d\eta^{A^{\prime}}.\nonumber\\
\end{eqnarray}
\normalsize
Using Eq. \eqref{eq:interchange}, we interchange the sum and the integration and then applying Eq. \eqref{eq:avg-state}, we get 
\begin{eqnarray}
&&\tilde{\rho}^{AB}_{\text{eff}} = \text{Tr}_{A'}  \sum_i \Big(\sqrt{\mathcal{M}_i^{A'A}}\otimes\mathbb{I}_2^B \Big)\times\nonumber\\
 && \Big( \int |\eta\rangle^{A^\prime}\langle\eta|   d\eta \Big)\otimes\tilde{\rho}^{AB}  \Big(\sqrt{(\mathcal{M}_i^{A'A})^{\dagger}}\otimes \mathbb{I}_2^B\Big)  \nonumber\\
&& = \text{Tr}_{A'}  \sum_i \Big(\sqrt{\mathcal{M}_i^{A'A}}\otimes\mathbb{I}_2^B \Big) \frac{1}{2}\mathbb{I}^{A'} \otimes \frac{1}{2}\mathbb{I}^{A} \otimes\nonumber\\ &&\Lambda^B  \Big(\sqrt{(\mathcal{M}_i^{A'A})^{\dagger}}\otimes \mathbb{I}_2^B \Big). 
\end{eqnarray}
When we perform weak Bell measurements with sharpness $\lambda$, $\tilde{\rho}^{AB}_{\text{eff}}$ becomes
\begin{equation}
\text{Tr}_{A'}  \sum_i \Big(\frac{x^2 + 2xy}{4}|B_i^{A'A}\rangle\langle B_i^{A'A}| + \frac{y^2}{4}\mathbb{I}_4^{A'A} \Big)\otimes \Lambda^B 
\end{equation}
with $x = \sqrt{\frac{1+3\lambda}{4}} - \sqrt{\frac{1-\lambda}{4}}$, and $y = \sqrt{\frac{1-\lambda}{4}}$. Now interchanging the sum and the trace, we obtain
\begin{eqnarray}
\tilde{\rho}^{AB}_{\text{eff}} &=& \sum_i \frac{(x+y)^2 + 3y^2}{4}  \Big(\frac{1}{2} \mathbb{I}_2^{A} \otimes \Lambda^B \Big) \nonumber \\
&=& \sum_i \frac{1}{4} \Big(\frac{1}{2} \mathbb{I}_2^{A} \otimes \Lambda^B \Big) \nonumber \\
&=& \frac{1}{2} \mathbb{I}_2^{A} \otimes \Lambda^B = \tilde{\rho}^{AB}.
\end{eqnarray}
Therefore, the effective state for the next round is same as the initial resource, and hence we obtain the proof.
\end{proof}

{\it Lemma 3}:
{\it When a Werner state of probability  $p'$ is used as the resource and a weak Bell measurement (Eq. \eqref{eq:whitenoise}), of sharpness $\lambda'$ is performed, the effective state, $\rho^{AB}(p',\lambda')$, shared between $A$ and $B$ for the next round turns out to be
\begin{equation}
\rho^{AB}(p',\lambda') = p'p(\lambda')|B_1^{AB}\rangle\langle B_1^{AB}|+ \frac{1-p'p(\lambda')}{4}\mathbb{I}_4^{AB},
\label{eq:rho-eff}
\end{equation}
where $p(\lambda')$ is the mixing parameter of the Werner state after the previous round.
}

\begin{proof}
When a weak Bell measurement of sharpness $\lambda'$ is employed, the effective state for a maximally entangled resource, $|B_1^{AB}\rangle$, is a Werner state with a probability, $p(\lambda')$, as in Eq. \eqref{eq:werner-phi}.  On the other hand, by substituting $\Lambda^B = \frac{1}{2}\mathbb{I}_2^B$ in Eq. \eqref{eq:product-resource} of lemma 2, we realize the effective state for the next round when  $\frac{1}{4}\mathbb{I}_4^{AB}$  is used as a resource is $\frac{1}{4}\mathbb{I}_4^{AB}$ itself. Therefore, when a Werner state of probability $p'$ is used as a resource and a weak Bell measurement of sharpness $\lambda'$ is performed, the effective state for the next round, by linearity is computed as 
\begin{eqnarray}
\rho^{AB}(p',\lambda')= p' \Big( p(\lambda')|B_1^{AB}\rangle\langle B_1^{AB}|+\nonumber\\ 
\frac{1-p(\lambda')}{4}\mathbb{I}_4^{AB} \Big) +
\frac{1-p'}{4}\mathbb{I}_4^{AB}\nonumber\\
=p'p(\lambda')|B_1^{AB}\rangle\langle B_1^{AB}|+ \frac{1-p'p(\lambda')}{4}\mathbb{I}_4^{AB}. 
\end{eqnarray}
 
Therefore, after this type of weak Bell measurement a Werner state remains an Werner state with updatation of mixing probability. 
\end{proof}


By recursion, we obtain that if the initial resource state is  $|B^{AB}_1\rangle$ and n-rounds of POVMs are performed on $A'A$ with sharpness values $\lambda_1, \lambda_2, \ldots, \lambda_{n-1}$, the resulting state for the i-th outcome takes the form as
\begin{eqnarray}
\rho_{AB}(p_n^{\lambda_1,\ldots,\lambda_{n-1}},\lambda_1, \lambda_2,\ldots,\lambda_{n-1})=\nonumber\\
p_n^{\lambda_1, \lambda_2,\ldots, \lambda_{n-1}} |B_i\rangle\langle B_i| + (1-p_n^{\lambda_1, \lambda_2,\ldots,\lambda_{n-1}})\frac{\mathbb{I}_4}{4}
\end{eqnarray}
with $p_n^{\lambda_1, \lambda_2, \ldots, \lambda_{n-1}}=p(\lambda_1)p(\lambda_2) \ldots p(\lambda_{n-1})$. The corresponding fidelity can be computed form lemma 1, given by $f(p_n^{\lambda_1,\ldots,\lambda_{n-1}},\lambda_1, \lambda_2,\ldots,\lambda_{n-1})=\frac{1}{2}(1+p_n^{\lambda_1,\ldots,\lambda_{n-1}})$. Applying these lemmas, we now state the following result.

{\it Theorem}:
{\it If a maximally entangled state is used as the initial resource,
\begin{equation}
\emph{MRN}_{f=2/3}(|B_1^{AB}\rangle) = 6,
\label{eq:mrn6}
\end{equation}
where $\emph{MRN}_{f=2/3}$ denotes the fact that $\emph{MRN}$ is computed for a fixed value of fidelity to be $2/3$.} 
\noindent Naturally, any other Bell state, $|B_i^{AB}\rangle$, would have the same MRN value. 

%
If the fidelity for each round is fixed to a value $f>\frac{2}{3}$, the maximal number of times, $\text{MRN}_f$, that one can continue the process, for an initial maximally entangled resource is given in Table \ref{tab:MRNf} (see Fig. \ref{fig:mrnvsf}).  Moreover, the effective state between $A$ and $B$ in the $(\text{MRN}_f + 1)-\text{th}$ round can still be entangled, i.e., can be used for teleportation with quantum fidelity lower than the prescribed fidelity $f$. We compute these ranges and note them down in Table \ref{tab:MRNf}.

\begin{table}[t]
\begin{center}
\begin{tabular}{|c|c|c|}
    \hline
  \hspace{0.3cm}  $\text{MRN}_f$ \hspace{0.3cm} & \hspace{0.3cm} Range of $f$ \hspace{0.3cm}  & \hspace{0.2cm} Range of $f_{ent}$  \hspace{0.2cm} \\
    \hline
    6 & 0.6666 - 0.6764 & -- \\
    \hline
    5 & 0.6765 - 0.6958 & 0.6765 - 0.6782 \\
    \hline
    4 & 0.6959 - 0.7227 & 0.6959 - 0.7025 \\
    \hline
    3 & 0.7228 - 0.7631 & 0.7228 - 0.7391 \\
    \hline
    2 & 0.7632 - 0.8333 & 0.7632 - 0.8028 \\
    \hline
    1 & 0.8334 - 1 & 0.8334 - 0.9553 \\
    \hline
   \end{tabular}
\caption{Maximal reattempt number ($\text{MRN}_f$) when fidelity at each round is fixed to be $f$. $f_{ent}$ denotes the  fidelity for a given $\text{MRN}_f$ such that the effective state for the next round, i.e., $(\text{MRN}_f + 1)^\text{th}$ round, is entangled but not enough to achieve the required fidelity $f$.}
\label{tab:MRNf}
\end{center}
\end{table}

Among the various ranges and critical values of $f$ obtained in Table \ref{tab:MRNf}, we want to highlight two points: Firstly, 
 for MRN$_f = 6$, the effective state for the next round is always unentangled. This is expected since the highest value of MRN is six (see Eq. \eqref{eq:mrn6}), and   existence  of any entangled effective state in the seventh round would imply MRN to be greater than six. Secondly, if the fidelity requirement at the first round is greater than $\frac{1}{6}(4+\sqrt{3})\approx 0.9553$, the state cannot be reused at all. This simply follows from the fact that the effective state for the second round becomes unentangled when the required $f > \frac{1}{6}(4+\sqrt{3})$. To obtain such  fidelity one has to perform a measurement with sharpness $\lambda > \frac{1}{3}(1+\sqrt{3}) \approx 0.9107$ resulting in a Werner state with $p \leq \frac{1}{3}$, an unentangled one (see Eq. \eqref{eq:p1}), as the effective state for the next round.

In the next section we would discuss other possible channels by which Bell measurements can be weakened. We show that among these possible weakening paths, the one considered in this section is the best for obtaining values of MRN.

\subsection{Weakening Bell measurements via local strategies}

Apart from the weakening strategies adopted in Eq. \eqref{eq:whitenoise}, there can be other ways by which Bell measurements can be weakened. We deal with some of these other weakening schemes in this section. Firstly, we consider POVMs by mixing Bell states with states from its ``Schmidt support'', i.e., the span of the Schmidt vectors of the same. 
 We find that, under such a scheme \emph{recycling the resource states for reattempting teleportation is not possible}. Then we consider cases which weakens the Bell measurements by mixing states from beyond its Schmidt support. Although in some cases, reattempts are possible, but among the considered weakening scenarios, the POVMs  in Eq.~\eqref{eq:whitenoise} gives the best MRN when a maximally entangled state is used as the initial resource state.

\subsubsection{Mixing states from the same Schmidt support}
When Bell states are weakened by mixing states from the same Schmidt support, we effectively get POVMs in the form $\{M_i^{A^{\prime}A}\}_{i=1}^4$ whose \(i\)-th element reads as
\begin{eqnarray}
\mathcal{M}_i^{A'A} = \mu |B_i^{A'A}\rangle\langle B_i^{A'A}| + \nonumber\\
(1-\mu)\hspace{0.1cm}\sigma_z^{A'}\otimes\mathbb{I}_2^A |B_i^{A'A}\rangle\langle B_i^{A'A}| \sigma_z^{A'}\otimes\mathbb{I}_2^A, 
\label{eq:colorednoise1}
\end{eqnarray}
where $\mu = \mu(\lambda)$ and $|B_i^{A'A}\rangle$ is one of the Bell states and $\sigma_i^{A'}(i=x, y, z)$ is the Pauli operator. These POVMs can be obtained by passing part of the Bell states through a single qubit local phase-flip  channel. Other examples of POVMs in a similar spirit include
\begin{eqnarray}
\lambda |B_{1,2}^{A'A}\rangle\langle B_{1,2}^{A'A}| +\nonumber\\ \frac{1-\lambda}{2} \Big(|00^{A'A}\rangle\langle 00^{A'A}|+ |11^{A'A}\rangle\langle 11^{A'A}|\Big), \nonumber\\
\lambda |B_{3,4}^{A'A}\rangle\langle B_{3,4}^{A'A}| +\nonumber\\ \frac{1-\lambda}{2} \Big(|01^{A'A}\rangle\langle 01^{A'A}|+ |10^{A'A}\rangle\langle 10^{A'A}|\Big), 
\end{eqnarray}
which reduce to Eq. \eqref{eq:colorednoise1} with $\mu=\frac{1+\lambda}{2}$. If we now use the strategy developed in the previous sections to investigate the fidelity and possibility of reusability of a maximally entangled resource subjected to weak measurements given in Eq. \eqref{eq:colorednoise1}, we find the  fidelity of the first round to be
\begin{equation}
f_{\text{phase-flip}}(\lambda) = \frac{1}{3}(1 + 2\lambda),
\label{eq:fid-phase-flip}
\end{equation}
where $f_{\text{phase-flip}}(\lambda)>\frac{2}{3}$ for $\lambda > \frac{1}{2}$. However, the effective state shared between Alice and Bob after the first round is diagonal and hence unentangled, which \emph{curbs the possibility of reusability of the resource state.} Thus,  weakening Bell measurements by mixing states from the same Schmidt support does not provide reusability even for a maximally entangled resource. In the next section, we consider other weakening schemes which in some cases enable reattempting teleportation for multiple rounds.

\subsubsection{Mixing states from the entire space}
Let us now  consider two distinct strategies where the Bell measurements are weakened by mixing states from different Schmidt supports, i.e., $\{B_1, B_2\}$ group elements are weakened by mixing states from the $\{B_3, B_4\}$ group and vice-versa. Such POVM elements can be obtained when a local bit-flip channel acts on $A'$.
The resulting i-th element of POVM takes the form
\begin{eqnarray}
\mathcal{M}_i^{A'A} = \lambda |B_i^{A'A}\rangle\langle B_i^{A'A}| + \nonumber\\  (1-\lambda)\hspace{0.1cm}\sigma_x^{A^{\prime}}\otimes\mathbb{I}_2^A |B_i^{A'A}\rangle\langle B_i^{A'A}| \sigma_x^{A^{\prime}}\otimes\mathbb{I}_2^A, 
\label{eq:colorednoise2a}
\end{eqnarray}
 Using these POVMs, the  fidelity obtained  after the first round with a maximally entangled resource (like the phase-flip case \eqref{eq:fid-phase-flip}) is given by
\begin{equation}
f_{\text{bit-flip}}(\lambda) = \frac{1}{3}(1 + 2\lambda),
\label{eq:fid-bit-flip}
\end{equation}
where $f_{\text{bit-flip}}(\lambda)>\frac{2}{3}$ for $\lambda > \frac{1}{2}$. The effective state shared between $A$ and $B$ for the second round  can be represented as
\begin{equation}
 \rho^{AB}_{\text{bit-flip}}(1,\lambda)= 
\begin{bmatrix}
   \frac{1}{4}       & 0 & 0  & \frac{\sqrt{\lambda(1-\lambda)}}{2} \\
    0       & \frac{1}{4} & \frac{\sqrt{\lambda(1-\lambda)}}{2} &  0 \\
     0   & \frac{\sqrt{\lambda(1-\lambda)}}{2}   & \frac{1}{4} & 0 \\
   \frac{\sqrt{\lambda(1-\lambda)}}{2}       & 0 & 0  & \frac{1}{4}
\end{bmatrix}.
\label{eq:effective-state-bitflip}
\end{equation}
Note that the entanglement of $\rho^{AB}_{\text{bit-flip}}(1,\lambda)$ as measured by concurrence \cite{conc1} reads as 
\begin{equation}
E\Big(\rho^{AB}_{\text{bit-flip}}(1,\lambda)\Big) = \max \Big\lbrace 0 ,  \sqrt{\lambda(1-\lambda)} - \frac{1}{2} \Big\rbrace,
\end{equation}
which is identically $0$ for $\lambda \in (0,1]$. Since the effective state for the second round is unentangled, the POVMs given in Eq. \eqref{eq:colorednoise2a} is ineffective for the reusability of the maximally entangled resource. 

The second  scheme involves  weakening the Bell measurements by mixing product states from orthogonal  compliment of the Schmidt support of the Bell state. For example, $|B_{1,2}\rangle$ and $|B_{3,4}\rangle$ are mixed with $(|01\rangle\langle 01|, |10\rangle\langle 10|)$ and $(|00\rangle\langle 00|, |11\rangle\langle 11|)$ respectively. The POVM elements read as
\begin{eqnarray}
\lambda |B_{1,2}^{A'A}\rangle\langle B_{1,2}^{A'A}| +\nonumber\\ 
\frac{1-\lambda}{2} \Big(|01^{A'A}\rangle\langle 01^{A'A}|+ |10^{A'A}\rangle\langle 10^{A'A}|\Big), \nonumber\\
\lambda |B_{3,4}^{A'A}\rangle\langle B_{3,4}^{A'A}| +\nonumber\\ 
\frac{1-\lambda}{2} \Big(|00^{A'A}\rangle\langle 00^{A'A}|+ |11^{A'A}\rangle\langle 11^{A'A}|\Big). \nonumber\\
\label{eq:povm-diff-support}
\end{eqnarray}
Using a maximally entangled state as resource, and employing the above measurements, the fidelity in the first round of teleportation is again computed to be
\begin{equation}
f_{\text{ortho-supp}}(\lambda) = \frac{1}{3}(1 + 2\lambda),
\label{eq:fid-ortho-supp}
\end{equation}
where $f_{\text{ortho-supp}}(\lambda)>\frac{2}{3}$ for $\lambda > \frac{1}{2}$. Although the expression of fidelity is same as the obtained for the POVM in Eq. \eqref{eq:colorednoise2a} the effective state after the first round in this case is entangled for some values of $\lambda$ and is computed to be
\begin{equation}
 \rho^{AB}_{\text{ortho-supp}}(1,\lambda)= 
\begin{bmatrix}
   \frac{1}{4}(2-\lambda)       & 0 & 0  & \sqrt{\frac{\lambda(1-\lambda)}{2}} \\
    0       & \frac{\lambda}{4} & 0 &  0 \\
     0   & 0   & \frac{\lambda}{4} & 0 \\
   \sqrt{\frac{\lambda(1-\lambda)}{2}}       & 0 & 0  & \frac{1}{4}(2-\lambda)
\end{bmatrix}.
\label{eq:effective-state-povm4}
\end{equation}
The entanglement of $\rho^{AB}_{\text{ortho-supp}}(1,\lambda)$ as measured by concurrence is computed to be 
\begin{equation}
E\Big( \rho^{AB}_{\text{ortho-supp}}(1,\lambda) \Big) = 2\max \Big\{ 0 , \sqrt{\frac{\lambda(1-\lambda)}{2}} - \frac{\lambda}{4} \Big\}.
\label{eq:ent-ortho-supp}
\end{equation}
In this case, the state now remains entangled for $\lambda < \frac{8}{9} \approx 0.8889$. 
Note that the effective state after the first round for the measurements (see Eq. \eqref{eq:whitenoise}), remains entangled for $\lambda < \frac{1}{3}(1+\sqrt{3}) \approx 0.9107$. 

\begin{figure}
\includegraphics[width=9cm]{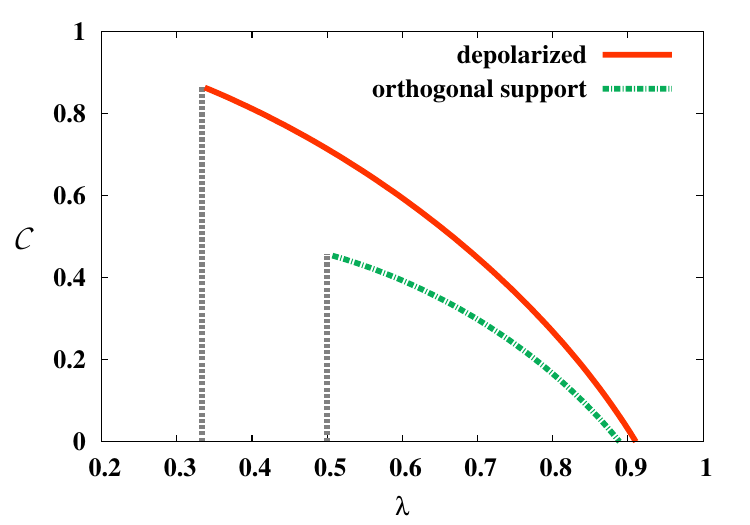}
\caption{Entanglement as quantified by concurrence, of the effective state after the first round, against the sharpness parameter of the measurement, \(\lambda\). 
Solid line corresponds to the measurement when single qubit of a Bell state is sent through a depolarization channel while the dashed one is when the Bell states are mixed with the product states from the orthogonal Schmidt support of the Bell state. In both the cases, maximally entangled state is the initial  resource. 
While the longitudinal axis is dimensionless, the vertical axis is in ebits.}
\label{fig:ent-com}
\end{figure}

For POVMs in Eq. \eqref{eq:whitenoise}, the region in the space of the sharpness parameter where the first round fidelities  are nonclassical and the effective state for the second round is entangled  is $\frac{1}{3} < \lambda < \frac{1}{3}(1+\sqrt{3})$, whereas the same for POVMs  in Eq. \eqref{eq:povm-diff-support} is for $\frac{1}{2} < \lambda < \frac{8	}{9}$. Therefore, the former measurement scheme leads to a higher reattempt number as well as a higher content of  entanglement than the one obtained by later procedure, as depicted in Fig. \ref{fig:ent-com}.


\section{Non-Maximally entangled state as initial resource}
\label{sec:nonmax}

In this section, instead of maximally entangled state as resource, we consider arbitrary pure state shared between $A$ and  $B$ under the same measurement strategies, discussed in Sec. \ref{sec:maxent}. We are now interested to investigate the change in the reusability number with the variation of entanglement content of the resource state.
Any pure bipartite state can always be written as
\begin{equation}
\label{puropure}
|\chi^{AB}\rangle=\sqrt{\alpha} |00\rangle+\sqrt{1-\alpha} |11\rangle,
\end{equation}
where $\alpha \in (0,1)$.
To teleport $|\eta\rangle^{A^\prime}$, the POVMs for weak Bell measurements described in Eq.~(\ref{eq:whitenoise}) is performed on the $A'A$ party of the input state $|\xi\rangle^{A'AB}=|\eta\rangle^{A'} \otimes |\chi^{AB}\rangle$. According to the protocol described in SubSec.~\ref{weak-me}, if $B$ wishes to complete the teleportation process by applying the proper unitaries based on the measurement outcomes, then
the corresponding  fidelity in the first round by averaging over all input states $|\eta\rangle^{A^\prime}$  is given by 
\begin{equation}
\label{dekh1}
f_\alpha(1,\lambda_1)=\frac{1}{2} (1-\lambda_1) +\frac{2}{3} \lambda_1 \Big[1+\sqrt{\alpha (1-\alpha)}\Big].
\end{equation}
Clearly, $f_\alpha(1,\lambda_1)>2/3$ for $\lambda_1 > \frac{1}{1+4 \sqrt{\alpha (1-\alpha)}}$.
However, if $B$ is indifferent and leaves the protocol, the effective state shared between $A$ and $B$  as a resource for the next (second) round computed according to Eq. \eqref{eq:rho-effective-gen} reads as
\begin{eqnarray}
\label{gen_eff1}
 \rho^{AB}_\alpha (1, \lambda_1) = p(\lambda_1)|\chi^{AB}\rangle\langle \chi^{AB}| + 
 \big(1-p(\lambda_1)\big) \times\nonumber\\
  \Big[ \frac{1}{2} \mathbb{I}^A_2 \otimes \Big(\alpha |0^B\rangle \langle 0^B| +
 (1-\alpha) |1^B\rangle \langle 1^B| \Big)  \Big],
\end{eqnarray}
where  $p(\lambda_1)$ is same as  in Eq. \eqref{eq:p1}.
Clearly, the effective state,  in Eq. (\ref{gen_eff1}), is an $X$ state, which reduces to a Werner state  for $\alpha=1/2$, i.e. for maximally entangled state.  
The  teleportation fidelity of an $X$ state of the form in Eq. \eqref{gen_eff1} after performing the weak Bell measurements is computed in the lemma stated below.

{\it Lemma 4}:
{\it The teleportation fidelity of an  $X$ state with probability $p'$,  
\begin{eqnarray}
\rho^{AB}_\alpha(p') = p' |\chi^{AB}\rangle\langle \chi^{AB}| &+&\nonumber\\
  \big(1-p'\big) \Big[ \frac{1}{2} \mathbb{I}^A_2 \otimes \Big(\alpha |0^B\rangle \langle 0^B| &+& (1-\alpha) |1^B\rangle \langle 1^B| \Big)  \Big],\nonumber\\
\end{eqnarray}
when subjected to a  weak POVMs in Eq. \eqref{eq:whitenoise} with sharpness parameter $\lambda'$, is given by
\begin{eqnarray}
f_\alpha(p',\lambda') = p' f_\alpha(1,\lambda') + \frac{1-p'}{2},
\label{gen_fid}
\end{eqnarray}
where the expression of $f_\alpha(1,\lambda')$ is expressed in Eq. \eqref{dekh1}.}
\begin{proof}
We proceed in a similar fashion as in lemma 1. The $|\chi\rangle^{AB}\langle \chi|$ part of $\rho^{AB}_\alpha(p')$ yields a fidelity of $f_\alpha(1,\lambda')$ on being subjected to weak Bell measurements with sharpness $\lambda'$. On the other hand, the separable part, $\frac{1}{2} \mathbb{I}^A_2 \otimes \big(\alpha |0\rangle^B \langle0|+(1-\alpha) |1\rangle^B \langle 1| \big)$, provides the fidelity of 
\begin{eqnarray}
\int_{|a|^2 + |b|^2 =1} && da \hspace{0.1cm} db \hspace{0.1cm} \big(|a|^2 \alpha + |b|^2 (1-\alpha)\big)  \nonumber \\
&& = \frac{1}{2}\alpha + \frac{1}{2}(1-\alpha) = \frac{1}{2}, 
\end{eqnarray}
where $a$ and $b$ are the coefficients of the arbitrary input state, $|\eta\rangle$ (see Eq. \eqref{eq:unknown_state}). Now, by using linearity, Eq.~(\ref{gen_fid}) is obtained.
\end{proof}

Like the Werner states, the $X$ states of the form obtained here, in our analysis, preserves its form when subjected to weak Bell measurements. We encapsulate this fact in the form of the following lemma.
 
{\it Lemma 5}:
{\it If an $X$ state with probability $p'$,  
\begin{eqnarray}
\label{x1}
\rho^{AB}_\alpha(p') = p' |\chi^{AB}\rangle\langle \chi^{AB}|&+&\nonumber\\
  \big(1-p' \big) \Big[ \frac{1}{2} \mathbb{I}^A_2 \otimes \big(\alpha |0^B\rangle \langle0^B|&+&(1-\alpha) |1^B\rangle \langle 1^B| \big)  \Big],\nonumber\\
\end{eqnarray}
 is used as the resource state shared between Alice and Bob and a weak Bell measurement as in Eq. \eqref{eq:whitenoise} of sharpness $\lambda'$ is performed, the effective state for the next round turns out to be an $X$ state of the same form with modified mixing parameter given by
\begin{eqnarray}
\rho^{AB}_{\alpha}(p',\lambda') = p'p(\lambda')|\chi^{AB}\rangle\langle \chi^{AB}|+\nonumber\\ 
 \Big(1-p'p(\lambda')\Big) 
  \Big[ \frac{1}{2} \mathbb{I}^A_2 \otimes\nonumber\\  \Big(\alpha |0^B\rangle \langle 0^B|+(1-\alpha) |1^B\rangle \langle 1^B| \Big)\Big].
 \label{x22}
\end{eqnarray}
}
\begin{proof}
By identifying $\frac{1}{2} \mathbb{I}^A_2 \otimes \big(\alpha |0^B\rangle \langle0^B|+(1-\alpha) |1^B\rangle \langle 1^B| \big)$ as $\frac{1}{2} \mathbb{I}^A_2 \otimes \Lambda^B$ with $\Lambda^B = \text{diag}(\alpha,1-\alpha)$,  it is evident from lemma 2 that an $X$ state of the form in Eq. \eqref{gen_eff1} will be an $X$ state of the same form after performing the weak Bell measurements. However, the change in probability of $X$ state for the next round, $p'p(\lambda')$,  follows from lemma 3.
\end{proof}

Using the above two lemmas, we can compute the fidelity and the effective state for any subsequent round when a pure non-maximally entangled bipartite state, Eq. (\ref{puropure}),  is shared between $A$ and $B$ as an initial resource, and POVMs in Eq. \eqref{eq:whitenoise}  is performed at each step.
Now we compute the $\text{MRN}_f$ for $|\chi\rangle^{AB}$  with  various values of $\alpha$ by fixing the  fidelity \(f\) at each round. See Table  \ref{tab:MRNfalpha} for the ranges of $\alpha$ which yield a given value of MRN. Obviously for $\alpha=1/2$, $|\chi^{AB}\rangle$ is  maximally entangled, and for that, we have already shown the $\text{MRN}_{f=2/3}$ to be six. The sharpness parameters  which provide  $f = 2/3$ in each round follows a recursion relation: 
\begin{equation}
\lambda_{i+1} = \frac{2 \lambda_i}{(1-\lambda_i)+\sqrt{(1-\lambda_i) (1+3 \lambda_i)}} = \frac{\lambda_i}{p(\lambda_i)},
\end{equation}
with $i=1,2,\ldots, 5$ and $\lambda_1 = \frac{1}{1+4 \sqrt{\alpha (1-\alpha)}} = \frac{1}{1+2\mathcal{C}_\alpha}$, where $\mathcal{C}_\alpha = 2\sqrt{\alpha (1-\alpha)}$ is the concurrence of the initial resource state. For a maximally entangled resource, the sharpness required to extract a first round fidelity of $2/3$ is given by $\lambda_1 = \frac{1}{3}$.

\begin{table}[t]
\begin{center}
\begin{tabular}{|c|c|c|}
    \hline
  \hspace{0.2cm}  $\text{MRN}_{f=2/3}$ \hspace{0.2cm} & \hspace{0.2cm} Range of $\alpha$ \hspace{0.3cm}  & \hspace{0.2cm} Range of $\alpha_{ent}$  \hspace{0.2cm} \\
    \hline
    6 & 0.5 - 0.3008 & -- \\
    \hline
    5 & 0.3007 - 0.1850 & 0.3007 - 0.2847 \\
    \hline
    4 & 0.1849 - 0.1087 & 0.1849 - 0.1606 \\
    \hline
    3 & 0.1086 - 0.0535 & 0.1086 - 0.0810 \\
    \hline
    2 & 0.0534 - 0.0159 & 0.0534 - 0.0273 \\
    \hline
    1 & 0.0158 - 0 & 0.0158 - 0.0007 \\
    \hline
   \end{tabular}
\caption{Maximal reattempt number ($\text{MRN}_f$) with respect to $\alpha$ when fidelity at each round is fixed to $f=2/3$. The second column gives the range of \(\alpha\) which gives $\text{MRN}_{f=2/3}$, while the third one mentions the range of \(\alpha\), denoted by \(\alpha_{ent}\) for which the output state is entangled.  }
\label{tab:MRNfalpha}
\end{center}
\end{table}

\begin{figure}[t]
\includegraphics[width=8.5cm]{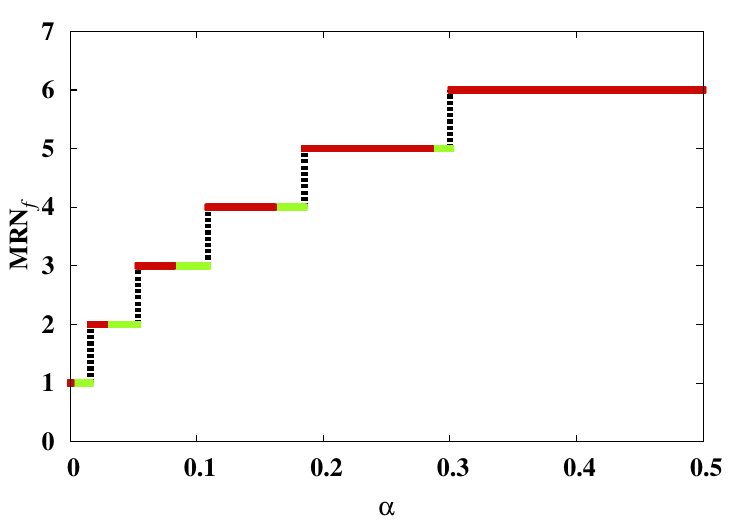}
\caption{$\text{MRN}_{f} \) against $\alpha\), the state parameter of a pure state.  The value of MRN$_f$ is computed by fixing the value of  fidelity, $f = \frac{2}{3}$, at every round. Both axes are dimensionless.}
\label{nonmax_fig}
\end{figure}
From our analysis with non-maximally entangled pure states as the initial resource, we stress two important facts: 

\emph{\(\emph{MRN} =6\) with non-maximally pure entangled states:} The most interesting scenario is that this margin ($\text{MRN}_{f=2/3}=6$) is preserved also  in the range $\alpha \in (0.3008,0.5]$ (see Table \ref{tab:MRNfalpha} and Fig. \ref{nonmax_fig}), which implies that  the non-maximally entangled states (of course within a particular range) is in the same footing with the maximally entangled one, in terms of the maximal number of  reattempts with the shared state. The maximal reattempt number decreases with the decreasing value of $\alpha$. See Table \ref{tab:MRNfalpha} for MRN$_f$ values with respect to $\alpha$.

\emph{Unutilized entanglement of $X$ states}:  In case of a non-maximally entangled initial resource, the effective state for the subsequent rounds are $X$ states for which the projective Bell measurements as well as the Pauli unitary operators are not optimal. Consequently, performing the Werner-type POVM, i.e. the weak Bell measurements in Eq.~(\ref{eq:whitenoise}) over the $X$ states only gives a bound to the maximal fidelity  for a given sharpness parameter. An upshot of the above analysis is reflected in the fact that for a non-maximally entangled initial resource, some entangled effective states in the next round, with  $\text{MRN}_f \leq 5$, do not yield quantum fidelities, i.e., cannot be used to increase MRN$_{f=2/3}$ following our strategy. However, some amount of entanglement  still exists  which can be used to obtain nonclassical fidelity in a teleportation scheme with other choices of measurements and unitaries. In Fig.~\ref{nonmax_fig}, the  dark (red) and grey (green) lines correspond to  the separable and entangled regions (measured by concurrence), respectively, for a certain range of $\alpha$.   Interestingly, we find that when $\text{MRN}=6$, all the effective states for different values of $\alpha$ become separable and hence are useless for teleportation.

\section{Conclusion}
\label{sec:conclusion}
Teleportation is one of the most fascinating inventions in  quantum theory. It has been experimentally verified and now with the satellite-based technology in the field, teleportation is marching fast in the direction of being realized on an intercontinental scale.

We addressed the issue of whether the resource state for teleportation can further be  used by another set of sender-receiver pair by 
using suitable  sets of  weak measurements, while maintaining nonclassical fidelities at every round of the use. 
 This recycling was achieved by not employing  complete projective measurements  which fully destroy the resource state after the very first use of the channel. 
 We observed that the sharpness of the measurements must be tuned in an appropriate manner so that it is weak enough to allow reattempts with the resource, yet adequately strong  to guarantee quantum fidelities during every use. 
 

We reported that if a maximally entangled state is  the initial resource, recycling the resource is possible  at most six times after applying the weak Bell measurements. Precisely, we compute the maximal reattempt number (MRN) for pure  maximally entangled  initial resource when the  fidelity at each round is just beyond  the classically achievable fidelity. Moreover, we found that among several weak measurement strategies, the one constructed by mixing Bell states with white noise led to a  higher value of MRN compared to other weakening measurement schemes. Interestingly, the MRN turned out to be six even for non-maximally entangled state having entanglement as measured by concurrence, higher than \(0.91\) ebits.  
We also studied the trends of MRN with the entanglement content of the resource state and the sharpness parameter of the measurement. 


The performance of quantum teleportation protocol was traditionally quantified via   single-shot  fidelity. Naturally, when reattempts are demanded, the scheme has to be redesigned to incorporate both MRN which involve measurement as well as state parameters and the  fidelity at every round for characterization. In this paper, we prescribe `a strategy' to meet both the demand for recycling the initial resource state and the same for nonclassical fidelities.
 Our work, therefore, opens up a new window of plausibility where for a fixed channel, one can extract quantum advantage in teleportation for several rounds and can address the trade-off   between information gain and disturbance due to measurement,  operationally.

%
%
%
%
%
%

\section{Acknowledgment}
S.R. acknowledges fruitful discussions with Arkaprabha Ghosal and Debarshi Das during QIPA 2018, organized by Harish Chandra Research Institute, Allahabad.
A.B. acknowledges the support of the Department of Science and Technology (DST), Government of India, through the award of an INSPIRE fellowship.


\begin{thebibliography}{99}
\bibitem{bd} The concurrence of a general Bell diagonal state, $\rho^{BD}$, with probabilities, $p_1, p_2, p_3, p_4$ in descending order is given by $C(\rho^{BD}) = \max \lbrace p_1-p_2-p_3-p_4, 0 \rbrace$. The usual Horodecki protocol yields a fidelity of $\frac{2+C(\rho^{BD})}{3}$, which is the maximal attainable fidelity for a given concurrence \cite{frank-ver}.
\bibitem{Bell-basis} The Bell basis consists of four maximally entangled states, given by
 \begin{eqnarray}
 |B_{i}^{AB}\rangle = \frac{1}{\sqrt{2}}(|00\rangle - (-1)^i |11\rangle),\,\, i =1,2, \nonumber \\
 |B_{i}^{AB}\rangle = \frac{1}{\sqrt{2}}(|01\rangle - (-1)^i |10\rangle), \,\, i=3, 4, \nonumber
\end{eqnarray} 
 where $\{|0\rangle,|1\rangle\}$ is the computational basis.
\end{thebibliography}
\end{document}